\documentclass[twocolumn]{openjournal}
\usepackage{natbib}
\usepackage{graphicx,amsmath,amssymb,amstext}
\usepackage{amsbsy,amsfonts,amsthm,color}
\usepackage{chemformula}
\usepackage[colorlinks,linkcolor=blue,citecolor=blue,urlcolor=blue ]{hyperref}
\usepackage[utf8]{inputenc}
\usepackage{float}
\usepackage[caption=false]{subfig}
\usepackage{booktabs}   
\usepackage{stfloats}

\newcommand{\kms}{\, {\rm km \, s^{-1}}}

\defcitealias{Prole2023}{LP23}
\defcitealias{Schauer2021}{AS21}

\begin{document}


\title{Assessing self-absorbed molecular lines as tracers of gravitational collapse\vspace{-3em}}

\author{Lillian Y. Cai$^{* \ 1,2}$}
\author{Felix D. Priestley$^{1}$}
\author{Sarah E. Ragan$^{1}$}
\email{$^*$email: lc5846@princeton.edu}

\affiliation {School of Physics and Astronomy, Cardiff University, Queen’s Buildings, The Parade, Cardiff CF24 3AA, UK}
\affiliation {Department of Astrophysical Sciences, Princeton University, Princeton, NJ, USA}

\date{July 2025}

\begin{abstract}

\noindent Red-shifted self-absorption features in molecular lines are commonly interpreted as signatures of gravitational collapse in pre- and protostellar cores. The shape of the line profile then encodes information on the dynamics of the collapse. There exist well-established observational techniques to estimate infall velocities from these profiles, but these have historically been calibrated on constant-velocity slab models, whereas more realistic simulations of gravitational collapse produce highly non-uniform radial velocity profiles. We produce synthetic line observations of a simulated collapsing prestellar core, including a treatment of the time-dependent chemical evolution. Applying observational techniques to the synthetic line profiles, we find that the estimated infall velocities are significantly and systematically lower than the mass-weighted infall velocities from the simulation. This is primarily because the self-absorption features tend to originate from the outer regions of the core, well beyond the location of the peak infall velocity. Velocities and mass accretion rates measured via these techniques are likely to underestimate the true values.

\end{abstract}


\maketitle

\section{Introduction} 
\label{sec:intro}
\noindent Dynamical evolution in molecular clouds takes place over timescales of millions of years \citep{bergin2007}, making it impossible to directly observe the collapse of material to stellar densities. A key indirect observational signature of star formation activity is `blue-asymmetric' profiles of optically-thick molecular lines, where the peak intensity is shifted bluewards of the rest velocity of the object in question \citep[e.g.][]{1998ApJ...504..900T,lee1999,lee2001}. These are typically interpreted as arising from red-shifted self-absorption: infalling material along the line of sight preferentially absorbing emission redwards of the rest velocity, as seen by an observer. While there are some caveats to this interpretation \citep[e.g.][]{smith2012}, these features remains widely used to identify regions of molecular clouds undergoing gravitational collapse \citep{campbell2016,redaelli2022}.

Beyond simply identifying that collapse is occurring, these self-absorbed line profiles can provide quantitative information as to the nature of this collapse. \citet{myers1996} demonstrated a simple analytical method to estimate infall velocities from line profile shapes, based on a model of two layers with different excitation temperatures and a velocity offset between them. By measuring five parameters related to the shape of the self-absorbed line profile, and the intrinsic velocity dispersion from an optically-thin line, the relative (i.e. infall) velocity between the layers can be obtained. A few further assumptions regarding the characteristic size and density of the emitting region provide the mass infall rate, which can be used to discriminate between competing theories of star formation \citep{maclow2004}.

While the simplicity of this model makes it attractive, it also necessarily limits its accuracy when applied to more complex situations. \citet{devries2005} previously investigated relaxing some of the assumptions of the original \citet{myers1996} model, in particular by allowing for varying excitation temperatures within the layers. However, their work still assumes that there is a single, constant velocity offset between the `foreground' and `background' layers. Even in the simplest cases of gravitational collapse in spherical symmetry, radial velocity profiles do not resemble this step-function picture; the velocity typically tends to zero at the centres and edges of simulations undergoing collapse, with a peak infall velocity at some intermediate radius depending on how far the collapse has proceeded \citep{larson1969,foster1993}. Moreover, different molecular lines will tend to probe different parts of this velocity profile, further complicating the situation \citep{smith2013,10.1093/mnras/stu1497,keown2016}.

In this paper, we combine hydrodynamical simulations with time-dependent chemistry and radiative transfer to produce synthetic line observations of model prestellar cores. We then obtain infall velocities from the line emission using established observational techniques, and assess how these values correpsond to the underlying velocities in the simulation. We find that measured infall velocities tend to underestimate the true mass-weighted value, often by a significant margin, and so should be treated with some caution.
\clearpage

\begin{table}[h]
  \centering
  \caption{Elemental abundances used in the chemical modelling.}
  \begin{tabular}{ccccc}
    \hline
    Element & Abundance & & Element & Abundance \\
    \hline
    H & $1.0$ & & He & $0.1$ \\
    C & $1.4 \times 10^{-4}$ & & S & $1.2 \times 10^{-5}$ \\
    N & $7.6 \times 10^{-5}$ & & Si & $1.5 \times 10^{-7}$ \\
    O & $3.2 \times 10^{-4}$ & & Mg & $1.4 \times 10^{-7}$ \\
    \hline
  \end{tabular}
  \label{tab:elem}
\end{table}

\begin{figure}[h]
	\includegraphics[width=\columnwidth]{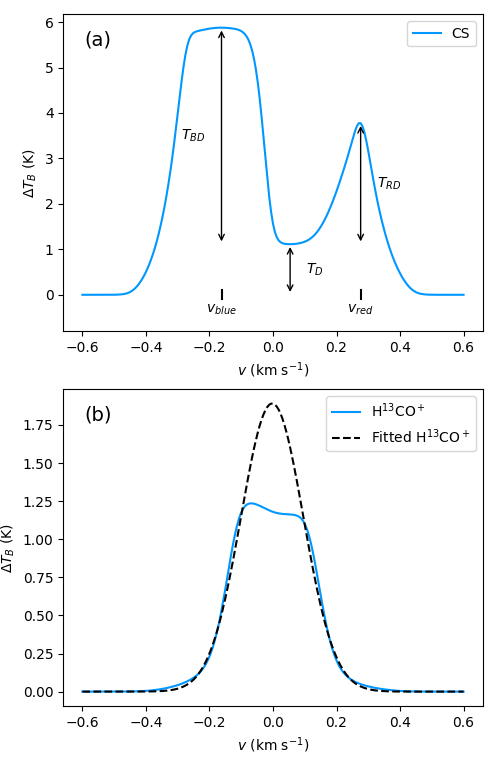}
	\\
	
\caption{Top: Line profile parameters in Eq.~\ref{eq:v_in}. Bottom: Example of fitting a Gaussian function to a self-absorbed line profile.}
\label{fig:gauss}
\end{figure}

\begin{table*}[!t]
\centering
\caption{Parameters required by Eq.~\ref{eq:v_in} to calculate $V_{in}$; red and blue peak velocities ($v_{red/blue}$) and height above dip ($T_{RD/BD}$), dip intensity ($T_D$), and bulk velocity dispersion ($\sigma_{bulk}$)}.
\label{tab:my-table}
\resizebox{\textwidth}{!}{%
\begin{tabular}{@{}cccccccccccc@{}}
\multicolumn{6}{c}{r = 0.05 pc, t = 0.81 Myr}                                       & \multicolumn{1}{l}{} & \multicolumn{5}{c}{r = 0.2 pc, t = 0.81 Myr}          \\ \cmidrule(r){1-6} \cmidrule(l){8-12} 
\multicolumn{1}{l}{Species} & $v_{red}$/km s$^{-1}$ & $v_{blue}$/km s$^{-1}$ & $T_D$/K & $T_{RD}$/K & $T_{BD}$/K &                      & $v_{red}$/km s$^{-1}$ & $v_{blue}$/km s$^{-1}$ & $T_D$/K & $T_{RD}$/K & $T_{BD}$/K \\ \cmidrule(r){1-6} \cmidrule(l){8-12} 
HCO$^+$ &
  \multicolumn{1}{l}{0.2460} &
  \multicolumn{1}{l}{-0.1740} &
  \multicolumn{1}{l}{2.2221} &
  \multicolumn{1}{l}{0.8509} &
  \multicolumn{1}{l}{4.5968} &
  \multicolumn{1}{l}{} &
  \multicolumn{1}{l}{0.1740} &
  \multicolumn{1}{l}{-0.1680} &
  \multicolumn{1}{l}{2.3182} &
  \multicolumn{1}{l}{0.0683} &
  \multicolumn{1}{l}{1.4714} \\
CS &
  \multicolumn{1}{l}{0.2280} &
  \multicolumn{1}{l}{-0.1500} &
  \multicolumn{1}{l}{1.1435} &
  \multicolumn{1}{l}{1.4731} &
  \multicolumn{1}{l}{4.5021} &
  \multicolumn{1}{l}{} &
  \multicolumn{1}{l}{0.0703} &
  \multicolumn{1}{l}{-0.0780} &
  \multicolumn{1}{l}{1.1492} &
  \multicolumn{1}{l}{-0.0934} &
  \multicolumn{1}{l}{1.4176} \\ \cmidrule(r){1-6} \cmidrule(l){8-12} 
Species                     & \multicolumn{5}{c}{Bulk Velocity ($\sigma_{bulk}$/km s$^{-1}$)}          & \multicolumn{1}{l}{} & \multicolumn{5}{c}{Bulk Velocity ($\sigma_{bulk}$/km s$^{-1}$)}          \\ \cmidrule(r){1-6} \cmidrule(l){8-12} 
N$_2$H$^+$                  & \multicolumn{5}{c}{0.1233}                            & \multicolumn{1}{l}{} & \multicolumn{5}{c}{0.0927}                            \\
H$^{13}$CO$^+$              & \multicolumn{5}{c}{0.1166}                            & \multicolumn{1}{l}{} & \multicolumn{5}{c}{0.0821}                            \\
C$^{18}$O                   & \multicolumn{5}{c}{0.1100}                            & \multicolumn{1}{l}{} & \multicolumn{5}{c}{0.0973}                            \\
C$^{34}$S                   & \multicolumn{5}{c}{0.1250}                            & \multicolumn{1}{l}{} & \multicolumn{5}{c}{0.1050}                            \\
\multicolumn{12}{l}{}                                                                                                                                              \\
\multicolumn{6}{c}{r = 0.05 pc, t = 0.87 Myr}                                       &                      & \multicolumn{5}{c}{r = 0.2 pc, t = 0.87 Myr}          \\ \cmidrule(r){1-6} \cmidrule(l){8-12} 
Species                     & $v_{red}$/km s$^{-1}$ & $v_{blue}$/km s$^{-1}$ & $T_D$/K & $T_{RD}$/K & $T_{BD}$/K &                      & $v_{red}$/km s$^{-1}$ & $v_{blue}$/km s$^{-1}$ & $T_D$/K & $T_{RD}$/K & $T_{BD}$/K \\ \cmidrule(r){1-6} \cmidrule(l){8-12} 
HCO$^+$                     & 0.2700    & -0.1800    & 2.2425 & 0.6304   & 4.3852   &                      & 0.1860    & -0.1780    & 2.2615 & 0.0174   & 1.2263   \\
CS                          & 0.2520    & 0.1560    & 1.1303 & 1.1572   & 4.2106   &                      & 0.0744    & -0.0660    & 1.1142 & -0.1037  & 1.1494   \\ \cmidrule(r){1-6} \cmidrule(l){8-12} 
Species                     & \multicolumn{5}{c}{Bulk Velocity ($\sigma_{bulk}$/km s$^{-1}$)}          &                      & \multicolumn{5}{c}{Bulk Velocity ($\sigma_{bulk}$/km s$^{-1}$)}          \\ \cmidrule(r){1-6} \cmidrule(l){8-12} 
N$_2$H$^+$                  & \multicolumn{5}{c}{0.1718}                            &                      & \multicolumn{5}{c}{0.1105}                            \\
H$^{13}$CO$^+$              & \multicolumn{5}{c}{0.1709}                            &                      & \multicolumn{5}{c}{0.0904}                            \\
C$^{18}$O                   & \multicolumn{5}{c}{0.1479}                            &                      & \multicolumn{5}{c}{0.1140}                            \\
C$^{34}$S                   & \multicolumn{5}{c}{0.1707}                            &                      & \multicolumn{5}{c}{0.1226}                           
\end{tabular}%
}
\end{table*}

\begin{figure*}[h]
    \centering
	\includegraphics[width=\columnwidth]{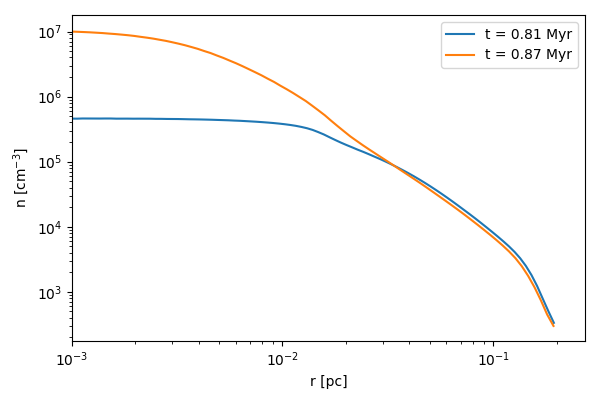}
	\includegraphics[width=\columnwidth]{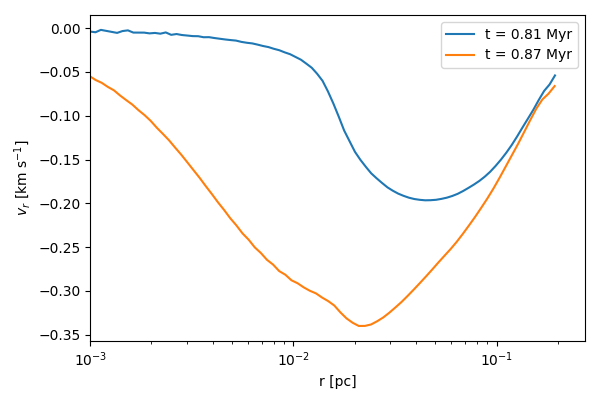}
	\\
	
\caption{Density (left) and radial velocity (right) profiles at t = $0.81\,\mathrm{Myr}$ and $0.87\,\mathrm{Myr}$.} 
\label{fig:dnv}
\end{figure*}

\begin{figure*}
    \centering
	\includegraphics[width=\columnwidth]{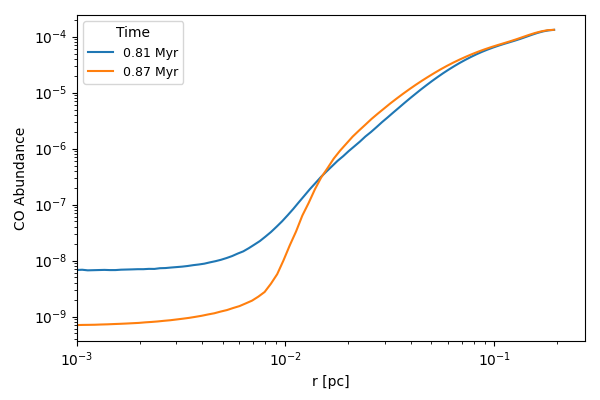}
	\includegraphics[width=\columnwidth]{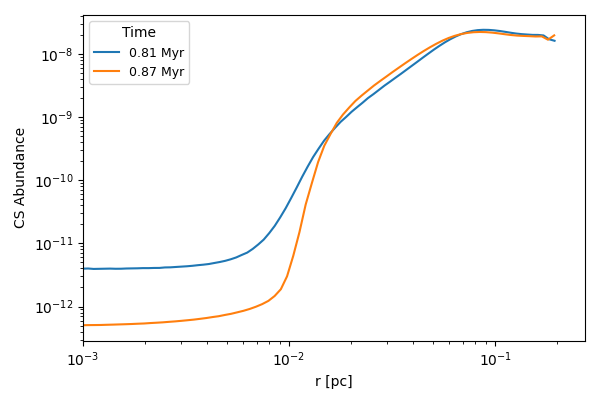}
	\includegraphics[width=\columnwidth]{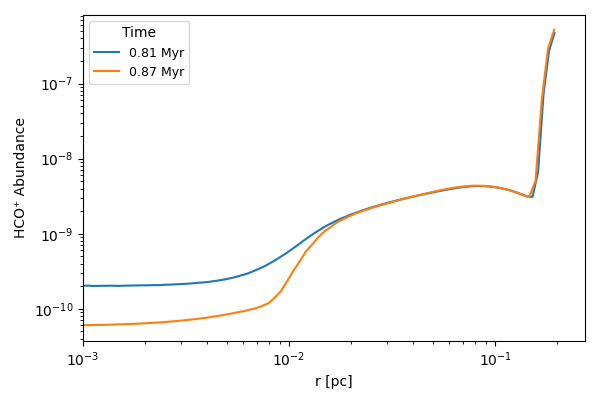}
	\includegraphics[width=\columnwidth]{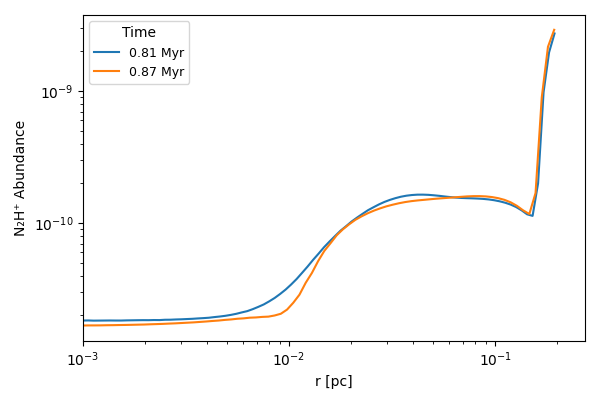}
	\\
	
\caption{Abundances of CO (top left), CS (top right), HCO$^+$ (bottom left), and N$_2$H$^+$ (bottom right) versus hydrogen nuclei density at t = $0.81\,\mathrm{Myr}$ and $0.87\,\mathrm{Myr}$.} 
\label{fig:abun}
\end{figure*}

\begin{figure*}
    \centering
	\includegraphics[width=\columnwidth]{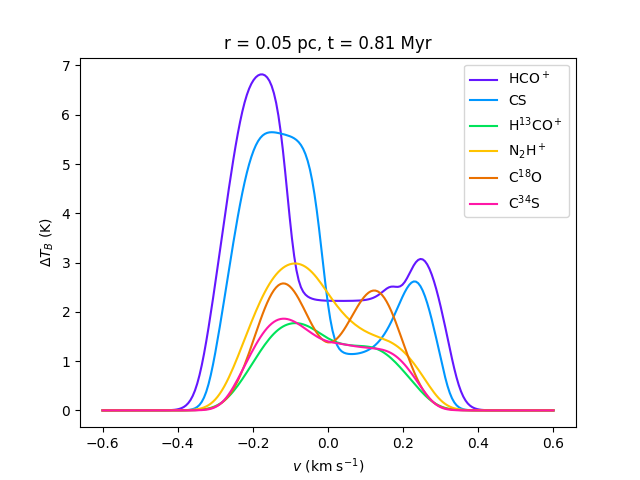}
	\includegraphics[width=\columnwidth]{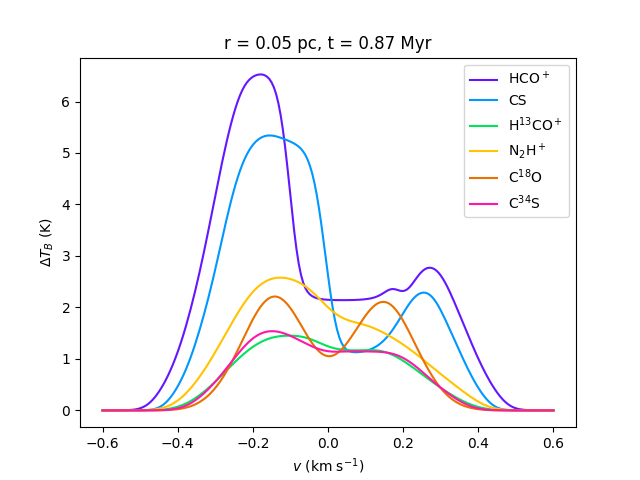}
	\includegraphics[width=\columnwidth]{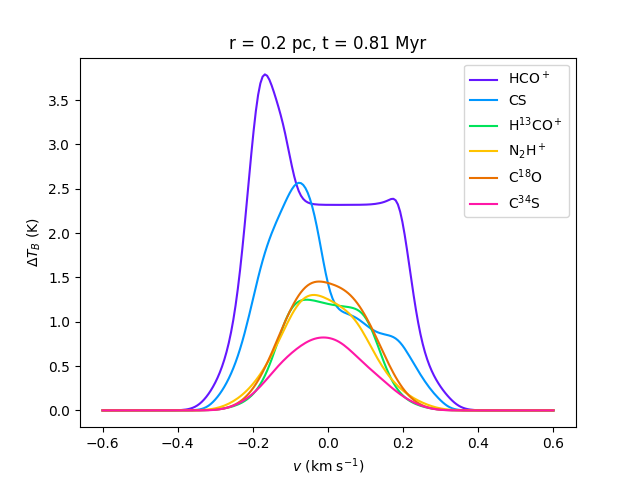}
	\includegraphics[width=\columnwidth]{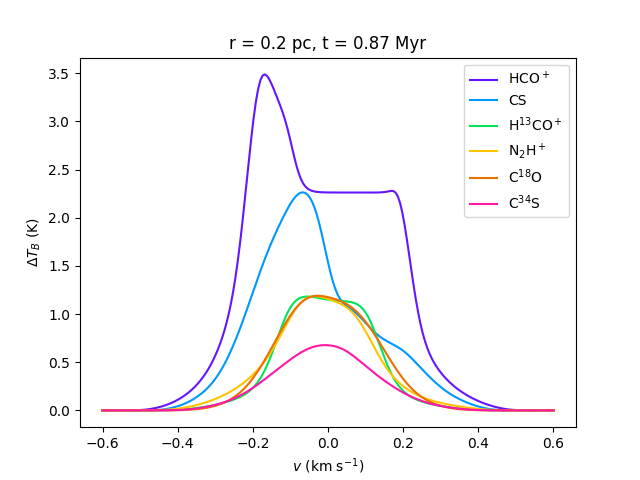}
	\\
	
\caption{Line profiles for four combinations of radii and timesteps. 
Top Left: r = $0.05 \, {\rm pc}$, t = $0.81\,\mathrm{Myr}$ - Top Right: r = $0.05 \, {\rm pc}$, t = $0.87\,\mathrm{Myr}$ - Bottom Left: r = $0.2 \, {\rm pc}$, t = $0.81\,\mathrm{Myr}$ - Bottom Right: r = $0.2 \, {\rm pc}$, t = $0.87\,\mathrm{Myr}$} 
\label{fig:line}
\end{figure*}

\begin{figure*}
    \centering
	\includegraphics[width=0.8\textwidth]{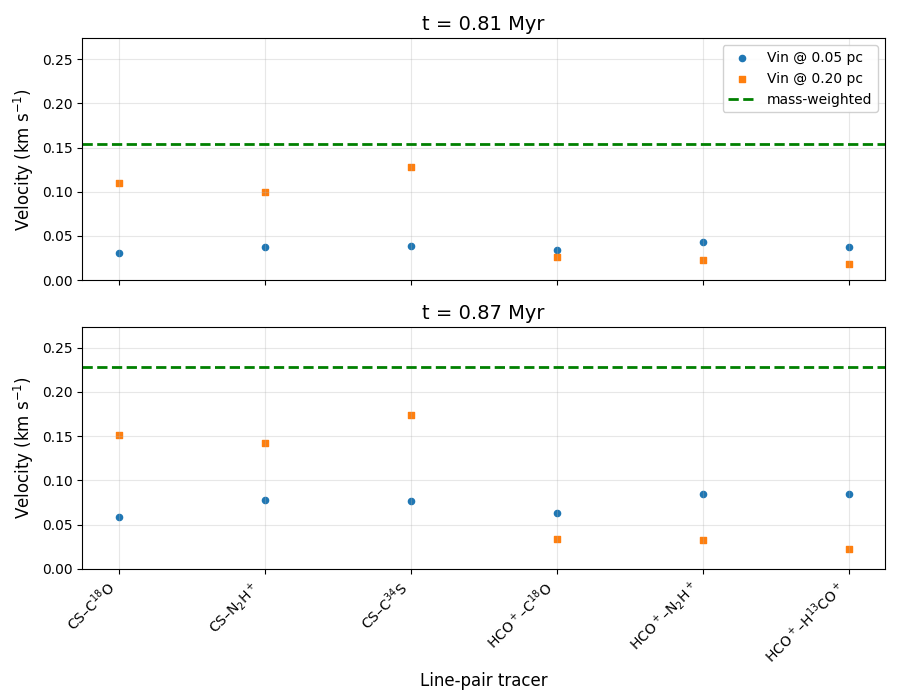}
	\\
	
\caption{True mass-weighted velocity and calculated infall velocities for all line-pair tracers, at t = $0.81\,\mathrm{Myr}$ and $0.87\,\mathrm{Myr}$.} 
\label{fig:result}
\end{figure*}

\begin{figure*}
    \centering
	\includegraphics[width=0.8\textwidth]{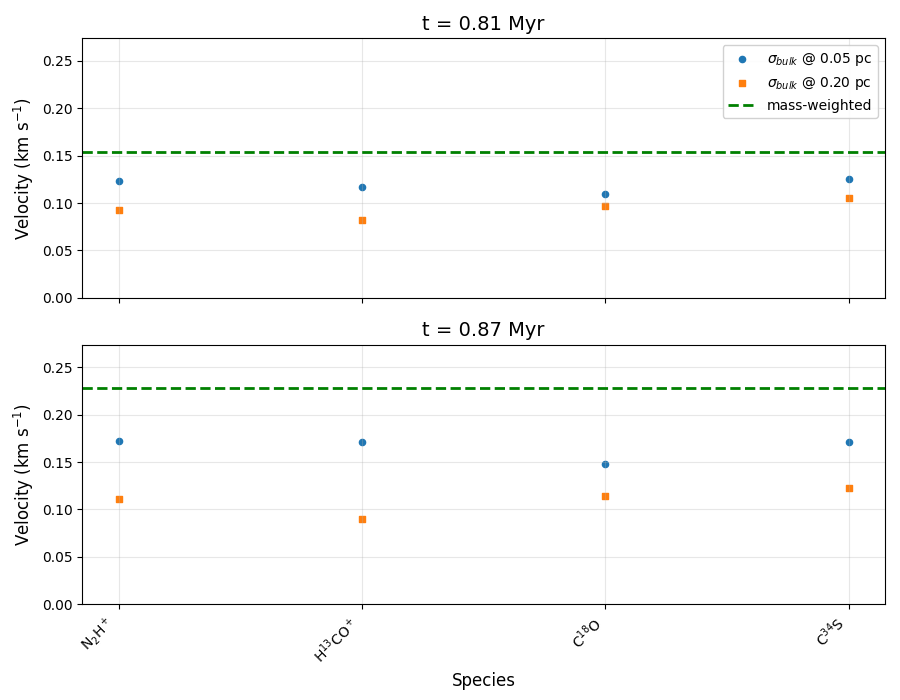}
	\\
	
\caption{True mass-weighted velocity and bulk velocities for the optically thin species, at t = $0.81\,\mathrm{Myr}$ and $0.87\,\mathrm{Myr}$.} 
\label{fig:result2}
\end{figure*}

\ \
\section{Method}
\label{sec:method}

We performed smoothed-particle hydrodynamic (SPH) simulations of a collapsing prestellar core using {\sc phantom} \citep{price2018}. Our initial conditions are a critical Bonnor-Ebert sphere \citep{bonnor1956,ebert1957} with mass $5 \, {\rm M_\odot}$, radius $0.221 \, {\rm pc}$ and sound speed $0.2 \, {\rm km \, s^{-1}}$ (i.e. molecular gas at $10 \, {\rm K}$), embedded in a background medium with a density $20$ times lower than the sphere's edge and sound speed raised to ensure pressure equilibrium. We use a particle mass of $2.7 \times 10^{-5} \, {\rm M_\odot}$, giving $\sim 2 \times 10^5$ particles within the sphere. We increase the sphere's mass by $10 \%$ to induce immediate collapse, and follow its evolution for $0.87 \, {\rm Myr}$, at which point the central density has reached $10^7 \, {\rm cm^{-3}}$.

We select all particles with a final radius within $0.2 \, {\rm pc}$ of the core's centre, and use their evolutionary histories as input for {\sc uclchem} \citep{holdship2017}, a time-dependent gas-grain chemical code, thereby obtaining self-consistent molecular abundance profiles rather than the constant values adopted in previous work. We use the UMIST12 \citep{mcelroy2013} gas-phase reaction network for species consisting of H, He, C, N, O, S, Si and Mg; the adopted elemental abundances are listed in Table \ref{tab:elem}. Freeze-out onto and desorption from grain surfaces are implemented following \citet{rawlings1992} and \citet{roberts2007} respectively. The cosmic ray ionisation rate is taken to be $1.3 \times 10^{-17} \, {\rm s^{-1}}$, and the external ultraviolet radiation field to be $1.7$ times the \citet{habing1968} field. We assume an external extinction of $A_{\rm V} = 3 \, {\rm mag}$ following \citet{aikawa2005}, and determine the internal extinction by integrating the radial density profile outward from each particle's position at each timestep.

We then performed radiative transfer modelling of the core's line emission using {\sc radmc3d} \citep{dullemond2012}. The SPH particle data are interpolated onto a $0.4 \, {\rm pc}$ Cartesian grid using {\sc splash} \citep{price2007}, and each cell is assigned the molecular abundances of the nearest post-processed SPH particle to the cell centre. Collisional data are taken from the LAMDA database \citep{schoier2005}. We model the $J=1-0$ rotational\footnote{We do not model the hyperfine structure of the N$_2$H$^+$ line; the single pure-rotational transition has a near-identical profile shape to the isolated hyperfine component most commonly used for kinematic studies \citep{priestley2023}.} transitions of C$^{18}$O, N$_2$H$^+$, HCO$^+$ and H$^{13}$CO$^+$, and the $J=2-1$ transitions of CS and C$^{34}$S; as the reaction network does not include the rarer isotopes ($^{13}$C, $^{18}$O and $^{34}$S), we scale the abundances of the main isotopologue by the isotopic ratios in \citet{wilson1994}, and assume identical collisional rates. Our output position-position-velocity cubes have spatial resolution $2 \times 10^{-3} \, {\rm pc}$ and velocity resolution $3 \times 10^{-3} \, {\rm km \, s^{-1}}$. We obtain line profiles by convolving the emission with a Gaussian `beam' of either $0.2$ or $0.05 \, {\rm pc}$ radius. One would resolve the 0.05 parsec scale with the IRAM-30m telescope when observing Orion B (distance = 400pc \citep{2017A&A...599A..98P}) in the 3mm lines (beam of 27’’ at 93 GHz). This is the primary single-dish facility with which such studies would be done.

Using data from the line profiles, we calculate the infall velocity for each combination of species. CS and HCO$^+$ are our optically thick self-absorbed lines, N$_2$H$^+$ and C$^{18}$O are optically thin, and C$^{34}$S and H$^{13}$CO$^+$ are to make sure our results aren't just due to chemical differences between optically thick and thin tracers. We find the infall velocities for apertures of radius $0.2$ and $0.05 \, {\rm pc}$, and timesteps $0.81\,\mathrm{Myr}$ and $0.87\,\mathrm{Myr}$. While the time interval is small in astronomical terms, this spans a period of rapid evolution in the model's density and velocity profiles. From \cite{myers1996}, we use

\begin{equation}
	\label{eq:v_in}
	V_{in} \approx \frac{\sigma_{bulk}^{2}}{v_{red} - v_{blue}} \ln\left(\frac{1 + eT_{BD}/T_D}{1 + eT_{RD}/T_D}\right),
\end{equation}

where $v_{red}$ and $v_{blue}$ are the velocities of the red and blue peaks of a self-absorbed line, $T_D$ is the intensity of the minimum of the self-absorption dip, and $T_{RD}$ and $T_{BD}$ are the heights of the red and blue peaks above this dip. A schematic illustration of these parameters is shown in Fig.~\ref{fig:gauss}$a$. $\sigma_{bulk}$ is the intrinsic non-thermal velocity dispersion, as obtained from a second, optically-thin line. 

From CS and HCO$^+$, which each have a prominent self-absorption dip, we can obtain the first four parameters using the blue and red peaks. N$_2$H$^+$ and C$^{18}$O, as well as C$^{34}$S and H$^{13}$CO$^+$, are more optically thin so they are used to obtain the non-thermal velocity dispersions. We fit a Gaussian distribution to their line profiles, masking the self-absorption dip if present, as illustrated in Fig.~\ref{fig:gauss}$b$. Each fitted curve has a standard deviation, which is taken as the observed velocity dispersion. We use this to calculate $\sigma_{bulk}$, the bulk velocity dispersion, using the equation $\sigma_{bulk}^{2} = \sigma_{obs}^{2} - \sigma_{therm}^{2}$, where $\sigma_{therm}^{2} = kT/m$, where $m$ is the molecular mass and $k$ is the Boltzmann constant. We set $T = 10K$, to obtain the $\sigma$ values that go into the infall velocity equation. Measured line parameters are listed in Table~\ref{tab:my-table}.

\section{Results}
\label{sec:results}

\noindent Fig.~\ref{fig:dnv} displays the radial velocity and density profiles at both $0.81\,\mathrm{Myr}$ and $0.87\,\mathrm{Myr}$. While the densities reach similar values at the sphere’s edge, about $3.0$--$3.3\times10^2\,\mathrm{cm}^{-3}$, the central density of the final timestep is over an order of magnitude larger than the central density at $0.81\,\mathrm{Myr}$, which levels off at $4.6\times10^5\,\mathrm{cm}^{-3}$. The radial velocity profiles both have a clear peak infall velocity, but the velocity profile at $0.87\,\mathrm{Myr}$ is noticeably broader and reaches a larger maximum by a factor of about $1.7$ compared to the $0.81\,\mathrm{Myr}$ curve. Fig.~\ref{fig:abun} shows that all 4 of the abundance profiles remain similar across both timesteps. The abundances decrease by at most one order of magnitude toward the center of the sphere, beginning at $r = 10^{-2}$ pc, due to freeze-out; however, this does not occur for N$_2$H$^+$\citep{2002ApJ...569..815T,2004A&A...416..191T}.

The sharp rise in the N$_2$H$^+$ and HCO$^+$ abundances beyond a radius of $0.15$ pc seen in Fig.~\ref{fig:abun} are an unphysical effect resulting from the low-density background medium in the simulation. At the timesteps where we analyse the simulation, the original `core' material has contracted to within $0.2$ pc, and been replaced by background-medium particles in the outer regions. This results in a steepening of the $r^{-2}$ density profile, visible in Fig.~\ref{fig:dnv}, and an increase in the abundances of the ionic species; these are primarily destroyed by dissociative recombination reactions, with rates scaling as $n^2$, and are thus more sensitive to the density than neutral species, where photodissociation reactions are more important. This behaviour does not occur in simulations which either lack a confining background medium \citep[e.g.][]{aikawa2005}, or do not feature a density discontinuity between the core and background \citep[e.g.][]{priestley2023a}. Although the increasing abundances are unphysical, they have a negligible impact on our results: the faster-than-$r^2$ decline of the density profile which causes the behaviour also means that very little mass is contained in this region, which therefore contributes minimally to the resulting line emission.

The line profiles for each of the radii and timestep combinations are shown in Fig.~\ref{fig:line}. Despite the physical differences between timesteps, the line profiles are fairly similar between $0.81\,\mathrm{Myr}$ and $0.87\,\mathrm{Myr}$, but appear to vastly differ for different aperture sizes. At r = $0.05 \, {\rm pc}$, each of the species C$^{34}$S, H$^{13}$CO$^+$, N$_2$H$^+$, and C$^{18}$O have much larger self-absorption dips, and the heights of the red and blue peaks of CS and HCO$^+$ are also nearly twice of those at r = $0.2 \, {\rm pc}$. In addition, the observed velocity dispersion is slightly larger at r = $0.05 \, {\rm pc}$, although only by approximately a factor of 2. 

Each of the infall velocities for the line-pair tracers, as well as the true mass-weighted velocities, are shown in Fig.~\ref{fig:result}. The infall velocities all fall significantly below the mass-weighted values of $0.153\,\mathrm{km\,s^{-1}}\;(t = 0.81\,\mathrm{Myr})$ and $0.228\,\mathrm{km\,s^{-1}}\;(t = 0.87\,\mathrm{Myr})$. Due to the similar line profile shapes, the \citet{myers1996} method gives similar values of $V_{in}$ for both timesteps, despite the later one having a significantly larger mass-weighted value. Using CS achieves results closest to the mass-weighted velocity, although only when using the larger of the two apertures. On the other hand, using HCO$^+$ often produces an infall velocity an entire order of magnitude smaller than the mass-weighted velocity. This discrepancy in results from using HCO$^+$ is due to the velocities at the red and blue peaks being much farther apart compared to those of CS. This increases $v_{red} - v_{blue}$ and produces a smaller $V_{in}$. The difference in results to those obtained from the CS line is despite the fact that the underlying velocity profile is identical, and arises purely from chemical and radiative transfer effects.

\section{Discussion}
\label{sec:discussion}
\noindent Our results in Fig.~\ref{fig:result} suggest that the \citet{myers1996} method may underestimate infall velocities in prestellar cores by up to an order of magnitude, depending on the choice of lines and the area over which they are measured. The application of the method to the prestellar core L1544 by \citet{myers1996} themselves returns a notably low infall velocity of $0.006 \kms$, consistent with our suggestion that it is likely an underestimate of the true value. As mass accretion rates estimated via this technique are linearly proportional to the measured $V_{in}$, they are subject to the same potentially-large source of error. While more sophisticated techniques have been developed \citep[e.g.][]{devries2005}, these are still calibrated using constant-velocity slab models, which we argue is the root cause of the discrepancy. For a two-layer model with a constant velocity, the velocity range of the self-absorption feature will naturally reflect this constant value. For more realistic structures, such as those in Fig.~\ref{fig:dnv}, there is no guarantee that the self-absorption feature (essentially corresponding to an optical depth of unity) will originate at a velocity corresponding to the mass-weighted infall velocity, or any other appropriate measure of the `average' infall rate. Lines from relatively-abundant molecules such as CS and HCO$^+$ will likely have self-absorption features primarily originating in the outer regions of the core, where velocities are relatively low, thus underestimating the actual rate of infall.

Modelling of self-absorbed line profiles is not the only available technique to determine infall velocities observationally; recent studies have often focused on directly measuring the velocity gradients of material accreting onto local centres of collapse \citep[e.g.][]{rigby2024,alvarez2024,sandoval2025}, exploiting the high sensitivities over large fields of view of modern millimetre/submillimetre facilities. This method comes with its own potential systematics, such as the unknown inclination of the accretion flows, and requires a much greater investment of telescope time than the single-pointing measurement required by the \citet{myers1996} method. As such, the latter still remains in common use, in combination with or as an alternative to methods based on velocity gradients \citep[e.g.][]{morii2025,moharana2026,beuther2026,gupta2026,xie2026}.

We note that the non-thermal velocity dispersion measured from optically-thin lines is, in these simulations, entirely caused by infall motions, and typically within a factor of two of the actual mass-weighted infall velocity (Fig.~\ref{fig:result2}). We suggest that in systems with red-shifted self-absorption features in optically-thick lines, the intrinsic velocity dispersion may be a better measure of the infall speed than values derived from line profile shapes, being largely unaffected by the issues discussed above. However, this interpretation may not hold in more complex, turbulent environments than the isolated collapsing cores considered here \citep{offner2008,smith2012,priestley2023,priestley2024}, and future work should explore whether \textit{any} observational signature is capable of probing the true mass-accretion rates of cores.

\section{Conclusion}
\label{sec:conclusion}
\noindent We have used synthetic observations of a simulated prestellar core to investigate the connections between observational infall diagnostics and the actual rate of collapse of the core. We find that commonly-used techniques based on constant-velocity, two-layer models of line emission invariably underestimate the true infall velocity of the simulation. This is because realistic, non-uniform velocity profiles peak in the inner regions of the core and tend to zero at the edge; self-absorption features generally originate from radii beyond that of the peak velocity, and so produce systematically lower infall rates than the true mass-weighted value. Prestellar mass accretion rates derived via these techniques are therefore likely underestimating the true values by potentially an order of magnitude.

\section*{Acknowledgements}
FDP and SER acknowledge the support of a consolidated grant (ST/W000830/1) from the UK Science and Technology Facilities Council (STFC).


\bibliographystyle{mnras} 
\bibliography{references, morereferences, citations3, citations4, citations5, citations6, citations7}
                                                                                                                                                                             \nocite{*}

\end{document}